\documentclass[%
superscriptaddress,
twocolumn,
aps,
pra,
]{revtex4-2}

\usepackage{graphicx}
\usepackage{dcolumn}
\usepackage{bm}
\usepackage{hyperref}
\hypersetup{colorlinks=true,citecolor=magenta,linkcolor=magenta,urlcolor=magenta}
\usepackage{braket}
\usepackage{xcolor}
\usepackage{colortbl}
\usepackage{amsmath,amssymb}
\usepackage[ruled,vlined]{algorithm2e}

\begin{document}
\preprint{APS/123-QED}

\title{Feedback-based quantum optimization and its classical counterpart: \\ quantum advantage and the power of classical algorithms}

\author{Tomohiro Hattori}
\affiliation{Graduate School of Science and Technology, Keio University, 3-14-1 Hiyoshi, Kohoku-ku, Yokohama-shi, Kanagawa 223-8522, Japan}
\affiliation{Basic Research Laboratories, NTT, Inc., Kanagawa 243-0198, Japan}
\author{Takuya Hatomura}
\affiliation{Basic Research Laboratories, NTT, Inc., Kanagawa 243-0198, Japan}
\affiliation{NTT Research Center for Theoretical Quantum Information, NTT, Inc., Kanagawa 243-0198, Japan}

\date{\today}
\begin{abstract}
Feedback-based quantum optimization is a quantum approach to combinatorial optimization. 
In this paper, we introduce the classical counterpart of feedback-based quantum optimization by using the quantum-classical correspondence of spin systems to discuss the possibility of quantum advantage. 
It also enables us to develop higher-order theory of a previously proposed classical approach to feedback-based quantum optimization. 
First, we compare the feedback-based algorithm for quantum optimization (FALQON) and its variant with their classical counterparts. 
Then, we perform benchmark tests of various quantum and classical algorithms with small-scale instances, and of classical algorithms with large-scale instances. 
Main findings are that (i) quantum algorithms can be advantageous to classical algorithms in terms of the quality of solutions, while classical algorithms tend to show faster convergence than quantum ones, and (ii) one of the classical algorithms discussed in this paper shows significant scalability for higher-order unconstrained binary optimization problems. 
These findings highlight the importance of quantumness and the usefulness of classical approaches. 
\end{abstract}

\keywords{}
\maketitle

\section{\label{sec:introduction}Introduction}

Combinatorial optimization is ubiquitous, with numerous applications across fields such as logistics~\cite{pillac2013review, golden2008vehicle, neukart2017traffic}, materials science~\cite{sanchez1984generalized, perdomo2012finding} and finance~\cite{fu1986application, yarkoni2022quantum}.
The importance and applications of solving combinatorial optimization problems have been broadened.
However, in computational complexity theory, most combinatorial optimization problems are classified as NP-hard, which indicates that there are no polynomial algorithms~\cite{garey1979computers}. 
In other words, exactly solving combinatorial optimization problems requires exponential computational cost with respect to the number of variables.

Combinatorial optimization problems can be formulated as quadratic unconstrained binary optimization (QUBO) forms, or equivalently, as the ground-state search in Ising spin-glass models~\cite {lucas2014ising}. 
Various physical systems have been studied as platforms for solving combinatorial optimization because of the latter formulation, and a wide range of solvers has been developed, including quantum annealers~\cite{johnson2011quantum} and quantum-inspired machines~\cite{yamaoka201520k,inagaki2016coherent,fixstars_amplify,goto2019combinatorial}. 
Numerous approaches have also been proposed under the QUBO (Ising) framework.
These include black-box optimization methods~\cite{FMA2020,FMA2022,photonic_laser2022,matsumori2022application,FMA2023,couzinie2025machine,tamura2026black,nakano2026swift}, large-scale combinatorial optimization~\cite{Karmi2017,Karimi2017_2,Irie2021Hybrid,Atobe_2022,kikuchi2023hybrid,hattori2025advantages,hattori2025impact,ide2025extending}, and constrained combinatorial optimization~\cite{hen2016quantum,hen2016driver,kudo2018constrained,kudo2020localization,hirama2023efficient,kanai2024annealing}.

Higher-order unconstrained binary optimization (HUBO) is higher-order formulation of QUBO and also has many applications~\cite{micheletti2021polymer, chermoshentsev2021polynomial, outeiral2021investigating, slongo2023quantum, jun2023hubo}.
To solve HUBO problems using QUBO solvers, higher-order interactions must be reduced to quadratic forms~\cite{biamonte2008nonperturbative,ishikawa2010transformation,kolmogorov2004energy,delong2012fast,freedman2005energy}. 
However, the proposed approaches typically require the introduction of auxiliary variables, leading to an increase in problem size and consequently to a degradation in performance.
Therefore, directly handling HUBO formulation is considered more efficient~\cite{ikeuchi2025evaluating}.

The feedback-based algorithm for quantum optimization (FALQON)~\cite{magann2022feedback,magann2022lyapunov} is one of the quantum algorithms for combinatorial optimization. 
A key idea of FALQON is the use of the Lyapunov control theory~\cite{cong2013survey} for reducing a cost function, and a main feature is the unnecessity of classical optimization unlike variational quantum algorithms~\cite{cerezo2021variational}. 
There are many studies based on FALQON, such as performance improvement~\cite{brady2025feedback, rattighieri2025accelerating, arai2025scalable,malla2024feedback,chandarana2024lyapunov,swain2025noise}, constrained combinatorial optimization~\cite{abdul2024feedback}, noise mitigation techniques~\cite{abdul2024adaptive}, and the ground-state preparation~\cite{larsen2024feedback, van2025imaginary}. 
One of the drawbacks is the substantial computational cost for iterative feedback due to the quantum version of Lyapunov control. 
A quantum iteration-free method using machine learning was proposed~\cite{perez2026learning}, but practical realization still costs heavily.

The counterdiabaticity-assisted classical algorithm for optimization (CACAO) is a specific case of FALQON that can be treated as a classical algorithm without substantial computational overhead~\cite{hatomura2025classical}.
This algorithm is derived from FALQON with the idea of local counterdiabatic driving~\cite{sels2017minimizing,PhysRevLett.123.090602,PhysRevA.103.012220,PhysRevB.106.155153,PhysRevX.14.011032,bhattacharjee2023lanczos,ohga2025improving,hatomura2024shortcuts}, which has the potential to improve quantum algorithms~\cite{PhysRevX.11.031070,Wurtz2022counterdiabaticity,PhysRevResearch.4.013141,malla2024feedback,chandarana2024lyapunov}.
CACAO exhibits faster convergence to low-energy states than FALQON, whereas it does not always converge to lower-energy states than FALQON. 
Although it is related to FALQON, it is not exactly the classical counterpart of FALQON.

In this paper, we introduce the classical counterpart of feedback-based quantum optimization by using the quantum-classical correspondence of spin systems and the canonical equations of motion for classical spins. 
This theory enables us to construct the exact classical counterparts of feedback-based quantum optimization algorithms and to discuss the possibility of quantum advantage. 
To be specific, we consider the classical counterparts of FALQON (CC-FALQON) and inhomogeneous FALQON (CC-iFALQON), where inhomogeneous FALQON (iFALQON) is a FALQON-like algorithm inspired by inhomogeneous driving for quantum annealing~\cite{susa2018quantum,susa2018expo}, and compare the performance of these algorithms. 
Moreover, this theory also enables us to develop the higher-order theory of CACAO (HOT-CACAO). 
In particular, we consider the introduction of the second-order counterdiabatic term to CACAO, which we simply call HOT-CACAO in this paper, and also introduce HOT-CACAO+, which includes all the terms we discuss in this paper. 
We perform benchmark tests of these algorithms with large instances of HUBO.

This paper is organized as follows. 
In Sec.~\ref{sec:quantum_algo}, we explain the feedback-based quantum optimization algorithms (FALQON and iFALQON).
In Sec.~\ref{sec:classical_algo}, we introduce the classical counterparts of the feedback-based quantum optimization algorithms (CC-FALQON and CC-iFALQON), and extend the classical algorithm (CACAO) to the higher-order classical algorithms (HOT-CACAO and HOT-CACAO+). 
In Sec.~\ref{sec:results}, we perform benchmark tests of the quantum and classical algorithms.
Further discussion is provided in Sec.~\ref{sec:discussion}.
We conclude the paper in Sec.~\ref{sec:conclusion}.

\section{Quantum algorithms\label{sec:quantum_algo}}
First, we explain the general idea of feedback-based quantum optimization, and then introduce the concrete algorithms. 
Finally, we mention the way of implementation. 
Throughout the paper, the Pauli matrices of $N$ quantum spins are expressed as $\{\hat{X}_i,\hat{Y}_i,\hat{Z}_i\}_{i=1}^N$. 

\subsection{Feedback-based quantum optimization}

In this section, we give a brief summary of feedback-based quantum optimization that uses the Lyapunov control theory proposed in Refs.~\cite{magann2022lyapunov, magann2022feedback}. 
We consider real-time quantum dynamics
\begin{equation}
    \label{eq:Seq}
    \mathrm{i}\frac{\partial}{\partial t}|\Psi(t)\rangle=\hat{H}(t)|\Psi(t)\rangle, 
\end{equation}
with the Hamiltonian in the following form:
\begin{equation}
    \label{eq:FQA}
    \hat{H}(t)=\beta_0(t)\hat{H}_\mathrm{P}+\sum_{k}\beta_k(t)\hat{V}_k. 
\end{equation}
Here, $\hat{H}_\mathrm{P}$ is the problem Hamiltonian consisting of the Pauli-Z operators and $\{\hat{V}_k\}_{k\neq0}$ are driving terms that do not commute with the problem Hamiltonian. 
When the parameters $\{\beta_k(t)\}_{k\neq0}$ are given by 
\begin{equation}
\label{eq:beta.ks}
    \beta_k(t)=\mathrm{i}\langle\Psi(t)|[\hat{H}_\mathrm{P},\hat{V}_k]|\Psi(t)\rangle,
\end{equation}
the cost function
\begin{align}
    \label{eq:Lyapunov}
    E_\mathrm{P}(t) =\bra{\Psi(t)}\hat{H}_\mathrm{P}\ket{\Psi(t)},
\end{align}
decreases with time evolution because $(d/dt)E_\mathrm{P}(t)\le0$ always holds. 
Note that $\beta_0(t)$ is an arbitrary function. 
Owing to the state-dependence of the parameters (\ref{eq:beta.ks}), this optimization must be performed in an iterative way, and thus it is called feedback-based quantum optimization.

\subsection{Algorithms\label{subsec:q_algo}}
We list the quantum algorithms discussed in this paper.

\subsubsection{FALQON\label{subsubsec:falqon}}

In the original proposal of FALQON~\cite{magann2022lyapunov, magann2022feedback}, the following Hamiltonian
\begin{equation}
\label{eq:falqon}
\hat{H}(t)=\hat{H}_\mathrm{P}+\beta^X(t)\sum_{i=1}^N\hat{X}_i,
\end{equation}
was considered. 
The single parameter $\beta^X(t)$ is given by
\begin{equation}
\label{eq:falqon_param}
\beta^X(t)=\mathrm{i}\langle\Psi(t)|\left[\hat{H}_\mathrm{P},\sum_{i=1}^N\hat{X}_i\right]|\Psi(t)\rangle, 
\end{equation}
which guarantees the reduction of the cost function (\ref{eq:Lyapunov}).

\subsubsection{iFALQON\label{subsubsec:iFALQON}}

As summarized above and mentioned in Refs.~\cite{magann2022lyapunov}, we can introduce multiple parameters.
For example, we can consider the following Hamiltonian
\begin{align}
    \label{eq:ifalqon}
    \hat{H}(t)&=\hat{H}_\mathrm{P}+\sum_{i=1}^N\beta_i^X(t)\hat{X}_i,
\end{align}
with the parameters
\begin{equation}
    \begin{aligned}
    \label{eq:ifalqon_param}
    \beta_i^X(t) = & \mathrm{i}\bra{\Psi(t)}[\hat{H}_\mathrm{P},\hat{X}_i]\ket{\Psi(t)}. 
\end{aligned}
\end{equation}
This algorithm is inspired by inhomogeneous driving for quantum annealing~\cite{susa2018quantum,susa2018expo}, and thus we call it inhomogeneous FALQON (iFALQON).

\subsection{Implementation}

\begin{algorithm}[t]
\caption{Experimental flow of feedback-based quantum optimization algorithms}
\label{alg:q_algorithm_hardware_flow}
\KwIn{Problem Hamiltonian $\hat{H}_\mathrm{P}$, driving terms $\{\hat{V}_k\}_{k\neq 0}$, initial state $\ket{\Psi(0)}$, operation time $T$, time step $\Delta t$}
\KwOut{Final solution 
$\{\langle\Psi(T)|\hat{Z}_i|\Psi(T)\rangle\}_{i=1}^N$ and final energy $E(T)$}
Initialize the control parameters $\{\beta_k(0)\}$ and the system Hamiltonian $\hat{H}(0)$\;
Set $t \gets \Delta t$\;
\While{$t \leq T$}{
    Prepare the initial state $\ket{\Psi(0)}$\;
    Evolve the system from $0$ to $t$ according to the system Hamiltonian scheduled in the previous steps\;
    Perform measurement, and obtain the control parameters $\{\beta(t)\}$ and the Hamiltonian $\hat{H}(t)$\;
    $t \gets t+\Delta t$\;
}
Prepare the initial state $\ket{\Psi(0)}$ once again\;
Evolve the system from $0$ to $T$ using the Hamiltonian scheduled in the previous steps\;
Perform measurement and obtain a solution $\{\langle\Psi(T)| \hat{Z}_i|\Psi(T)\rangle\}_{i=1}^N$ and the final energy estimate $E(T)$\;
\end{algorithm}

The experimental procedure of feedback-based quantum optimization algorithms is shown in Algorithm~\ref{alg:q_algorithm_hardware_flow}.
As input, the problem Hamiltonian $\hat{H}_\mathrm{P}$ is given and the system is initialized in an initial state $\ket{\Psi(0)}$.
The control parameters and the system Hamiltonian are also initialized as $\{\beta_k(0)\}$ and $\hat{H}(0)$ from the input.
The protocol proceeds iteratively until the total time $T$ is reached. 
At each iteration step, the system is first prepared in the initial state, and then time-evolved from $t=0$ to $t=n\Delta t$ for the $n$th iteration step ($\Delta t$ is a certain time step) under the Hamiltonian scheduled in the previous iterations. 
Measurement determines the control parameters $\{\beta_k(n\Delta t)\}$ and the Hamiltonian $\hat{H}(n\Delta t)$ for the next iteration step. 
After iterations, the initial state is prepared again and evolved from $t=0$ to $t=T$ using the scheduled Hamiltonian. 
Finally, we obtain a solution $\{\langle\Psi(T)|\hat{Z}_i|\Psi(T)\rangle\}_{i=1}^N$ with an energy estimate $E(T)$.

In our numerical calculation, the initial state is the uniform superposition over the computational basis, which corresponds to the ground state of the transverse-field Hamiltonian $\hat{V}=-\sum_{i=1}^N\hat{X}_i$.
Results are obtained by numerically solving the Schr\"{o}dinger equation with QuTiP~\cite{QuTiP1, QuTiP2}.
Thus, we numerically obtain a quantum state at each time without destroying it.

\section{Classical algorithms\label{sec:classical_algo}}
First, we derive the classical counterpart of feedback-based quantum optimization, and then introduce the concrete algorithms. 
Finally, we explain the way of implementation. 
Throughout the paper, $N$ classical spins are given by unit vectors $\{m_i^X,m_i^Y,m_i^Z\}_{i=1}^N$, where $|\bm{m}_i|=1$ for each classical spin vector $\bm{m}_i=(m_i^X,m_i^Y,m_i^Z)$.

\subsection{Classical counterpart of feedback-based quantum optimization\label{sec:classical.theory}}

In this section, we derive the classical counterpart of feedback-based quantum optimization. 
By replacing the Pauli matrices $\{\hat{X}_i,\hat{Y}_i,\hat{Z}_i\}_{i=1}^N$ with the classical spins $\{m_i^X,m_i^Y,m_i^Z\}_{i=1}^N$, we can obtain the classical counterpart of a given quantum-spin Hamiltonian. 
The canonical equations of motion for these classical spins are given by
\begin{equation}
\begin{aligned}
\label{eq:equation_of_motion}
    &\frac{d}{dt}\bm{m}_i=2\bm{m}_i\times\bm{h}_i^\mathrm{eff},\\
    &\bm{h}_i^\mathrm{eff}=-\frac{\partial\mathcal{H}_t}{\partial\bm{m}_i},
\end{aligned}
\end{equation}
where $\mathcal{H}_t$ is the classical-spin Hamiltonian obtained by the above replacement. 
These canonical equations of motion correspond to the Schr\"odinger dynamics (\ref{eq:Seq}) of the quantum counterpart (see Ref.~\cite{hatomura2018shortcuts} or Appendix~\ref{appendix:quantum_classical_correspondence}).

The classical counterpart of the problem Hamiltonian $\mathcal{H}_\mathrm{P}$ can be directly regarded as the classical counterpart of the cost function (\ref{eq:Lyapunov}), i.e., $E_\mathrm{P}(t)=\mathcal{H}_\mathrm{P}$, and its time derivative is given by
\begin{equation}
\begin{aligned}
\label{eq:time_derivative_equation_of_motion}
\frac{d}{dt}\mathcal{H}_\mathrm{P}&=\sum_{i=1}^N\frac{dm_i^Z}{dt}\frac{\partial\mathcal{H}_\mathrm{P}}{\partial m_i^Z}\\
&=-2\sum_{k\neq0}\beta_k(t)\sum_{i=1}^N\left(m_i^X\frac{\partial\mathcal{V}_k}{\partial m_i^Y}-m_i^Y\frac{\partial\mathcal{V}_k}{\partial m_i^X}\right)\frac{\partial\mathcal{H}_\mathrm{P}}{\partial m_i^Z},
\end{aligned}
\end{equation}
where $\mathcal{V}_k$ is the classical counterpart of $\hat{V}_k$. 
Thus, the cost function decreases when we set
\begin{equation}
\beta_k(t)=2\sum_{i=1}^N\left(m_i^X\frac{\partial\mathcal{V}_k}{\partial m_i^Y}-m_i^Y\frac{\partial\mathcal{V}_k}{\partial m_i^X}\right)\frac{\partial\mathcal{H}_\mathrm{P}}{\partial m_i^Z},\quad\text{for }k\neq0.
\end{equation}
Note that $\beta_0(t)$ is again an arbitrary function. 
Remarkably, the classical counterpart of feedback-based quantum optimization does not require any iterative process.

\subsection{Algorithms\label{subsec:c_algo}}
We list the classical algorithms discussed in this paper.

\subsubsection{CC-FALQON\label{subsubssec:falcon}}
We introduce the classical counterpart of FALQON (CC-FALQON).
In CC-FALQON, we consider the following Hamiltonian
\begin{equation}
\begin{aligned}
\label{eq:ccfalqon}
&\mathcal{H}_t=\mathcal{H}_\mathrm{P}+\beta^X(t)\sum_{i=1}^Nm_i^X,
\end{aligned}
\end{equation}
with the parameter
\begin{equation}
\label{eq:ccfalqon_param}
\beta^X(t)=-2\sum_{i=1}^{N}m_i^Y\frac{\partial\mathcal{H}_\mathrm{P}}{\partial m_i^Z}.
\end{equation}
This is the exact classical counterpart of FALQON.

\subsubsection{CC-iFALQON\label{subsubssec:falcon-md}}
We also introduce the classical counterpart of iFALQON (CC-iFALQON). 
In CC-iFALQON, we consider the following Hamiltonian
\begin{equation}
\begin{aligned}
\label{eq:ccifalqon}
&\mathcal{H}_t=\mathcal{H}_\mathrm{P}+\sum_{i=1}^N\beta_i^X(t)m_i^X,
\end{aligned}
\end{equation}
with the parameters
\begin{equation}
\label{eq:ccifalqon_param}
\beta_i^X(t)=-2m_i^Y\frac{\partial\mathcal{H}_\mathrm{P}}{\partial m_i^Z}.
\end{equation}
This is the exact classical counterpart of iFALQON. 

\subsubsection{CACAO\label{subsubsec:cacao}}
In CACAO~\cite{hatomura2025classical}, which was derived from the theory of FALQON, we consider the following Hamiltonian
\begin{equation}
\begin{aligned}
\label{eq:cacao}
&\mathcal{H}_t=\sum_{i=1}^N\beta_i^Y(t)m_i^Y,
\end{aligned}
\end{equation}
with the parameters
\begin{equation}
\label{eq:cacao_param}
\beta_i^Y(t)=2m_i^X\frac{\partial\mathcal{H}_\mathrm{P}}{\partial m_i^Z}.
\end{equation} 
In the previous study~\cite{hatomura2025classical}, faster convergence than that of FALQON was confirmed, whereas the quality of solutions was worse than that of FALQON.

\subsubsection{HOT-CACAO\label{subsubsec:hcacao}}
The general idea introduced in Sec.~\ref{sec:classical.theory} enables to develop the higher-order theory of CACAO (HOT-CACAO).
For example, HOT-CACAO with the second-order counterdiabatic term uses the following Hamiltonian
\begin{equation}
\label{eq:hcacao}
\mathcal{H}_t 
= \sum_{i=1}^{N}\beta_i^Y(t)m_i^Y + \sum_{\substack{i,j=1 \\ (i\neq j)}}^{N}\beta_{ij}(t)(m_i^Ym_j^Z+m_i^Zm_j^Y),
\end{equation}
with the parameters
\begin{equation}
\label{eq:hcacao_param}
\begin{aligned}
&\beta_i^Y(t)=2m_i^X\frac{\partial\mathcal{H}_\mathrm{P}}{\partial m_i^Z},\\
&\beta_{ij}(t) = 2m_i^Xm_j^Z\frac{\partial\mathcal{H}_\mathrm{P}}{\partial m_i^Z}.
\end{aligned}
\end{equation}
Note that HOT-CACAO is not limited to the second-order one theoretically.

\subsubsection{HOT-CACAO+\label{subsubsec:hcacao-x}}
Finally, we introduce HOT-CACAO+, which includes all the terms discussed in this paper. 
The Hamiltonian is given by
\begin{equation}
\begin{aligned}
\label{eq:hcacao+}
\mathcal{H}_t=&\mathcal{H}_\mathrm{P}+ \sum_{W=\{X,Y\}}\sum_{i=1}^{N}\beta_i^W(t)m_i^W\\
&+ \sum_{\substack{i,j=1 \\ (i\neq j)}}^N\beta_{ij}(t)(m_i^Ym_j^Z+m_i^Zm_j^Y),
\end{aligned}
\end{equation}
with the parameters 
\begin{equation}
\label{eq:hcacao+_param}
\begin{aligned}
&\beta_i^X(t)=-2m_i^Y\frac{\partial\mathcal{H}_\mathrm{P}}{\partial m_i^Z},\\
&\beta_i^Y(t)=2m_i^X\frac{\partial\mathcal{H}_\mathrm{P}}{\partial m_i^Z},\\
&\beta_{ij}(t) = 2m_i^Xm_j^Z\frac{\partial\mathcal{H}_\mathrm{P}}{\partial m_i^Z}.
\end{aligned}
\end{equation}
This algorithm has the highest expressionability in the parameter space discussed in this paper.

\subsection{Implementation\label{subsec:settings_classical}}

\begin{algorithm}[t]
\label{alg:algorithm_flow}
\caption{Computational flow of classical optimization algorithms}
\KwIn{Problem Hamiltonian $\mathcal{H}_\mathrm{P}$, driving terms $\{\mathcal{V}_k\}$, initial state $\{\bm{m}_i(0)\}$, operation time $T$, time step $\Delta t$}
\KwOut{Final configuration $\{\bm{m}_i(T)\}_{i=1}^N$ and energy $\mathcal{H}_P|_{t=T}$}
Initialize the control parameters $\{\beta_k(0)\}$ and the system Hamiltonian $\mathcal{H}_{t=0}$\;
$t\gets0$\;
\While{$t<T$}{
    Solve the equations of motion and obtain $\{\bm{m}_i(t+\Delta t)\}_{i=1}^N$\;
    Calculate the parameters $\{\beta_k(t+\Delta t)\}$ and the system Hamiltonian $\mathcal{H}_{t+\Delta t}$\;
    $t\gets t+\Delta t$\;
}
Obtain the solution $\{\bm{m}_i(T)\}_{i=1}^N$ and the energy estimate $\mathcal{H}_P|_{t=T}$\;
\end{algorithm}
The procedure of classical algorithms is shown in Algorithm~\ref{alg:algorithm_flow}. 
As input, the problem Hamiltonian $\mathcal{H}_\mathrm{P}$ and the initial state $\{\bm{m}_i(0)\}_{i=1}^N$ are given. 
The initial parameters $\{\beta_k(0)\}$ and the system Hamiltonian $\mathcal{H}_{t=0}$ are calculated. 
Then, we solve the equation of motion from $t=0$ to $t=T$. 
At each numerical time step $\Delta t$, the control parameters $\{\beta_k(t)\}$ are calculated with the current state of the system. 
After completing the time evolution, the final spin configuration $\{\bm{m}_i(T)\}_{i=1}^N$ is obtained, and the corresponding energy $\mathcal{H}_\mathrm{P}|_{t=T}$ is evaluated.

The computational cost of CACAO for QUBO was shown as $O(N^2T)$ in the worst case~\cite{hatomura2025classical}.
In CC-FALQON, CC-iFALQON, HOT-CACAO, and HOT-CACAO+, the computational cost increases as $O(N^3T)$ in the worst case since we have to consider additional interaction terms at each step.
Note that we can reduce computational cost in the case of a sparse graph as $O(dNT)$ in CACAO~\cite{hatomura2025classical}, $O(d^2NT)$ in CC-FALQON and CC-iFALQON, and $O(dN^2T)$ in HOT-CACAO and HOT-CACAO+ ($O(d^2NT)$ if the second-order counterdiabatic term is introduced only to the same graph topology of the problem Hamiltonian) with the degree of the vertices $d$. 
For HUBO, we need additional computational cost for higher-order interactions.

In these classical algorithms, we set the initial state in two ways: a fixed initial state and a random initial state.
We set a fixed initial state as  $(m_i^X(0), m_i^Y(0), m_i^Z(0)) = (1, 0, 0)$, which corresponds to the ground state of the transverse-field Hamiltonian $\mathcal{V}=-\sum_{i=1}^Nm_i^X$. 
The random initial states are randomly generated as $(m_i^X(0), m_i^Y(0), m_i^Z(0)) = (\sin\theta \cos\phi, \sin\theta \sin\phi, \cos\theta)$, where $\theta$ and $\phi$ are sampled uniformly from the intervals $[0,\pi]$ and $[0,2\pi)$, respectively. 
For CACAO, however, we set $\phi = 0$ because the dynamics is restricted to the $X$--$Z$ plane.
For HOT-CACAO, we also set $\phi=0$ because the dynamics is simplified to the $X$--$Z$ plane when we set $m_i^Y(0)=0$.
The equations of motion are solved by the fourth-order Runge-Kutta method.

\section{Benchmark tests\label{sec:results}}
We perform benchmark tests of the proposed algorithms on representative combinatorial optimization problems. 
First, we compare the quantum algorithms with the classical counterparts to discuss the possibility of quantum advantage. 
Then, we assess the performance of all the algorithms in terms of convergence behavior and solution quality.
We then extend the analysis of classical algorithms to large-scale and higher-order optimization problems, examining their performance.

We consider the $k$-SAT problem to evaluate their performance on both QUBO and HUBO forms.
The $k$-SAT problem is a Boolean satisfiability problem defined on a set of Boolean variables.
An instance consists of a logical formula composed of clauses, where each clause is a disjunction of $k$ literals.
A literal corresponds to either a variable or its negation.
The clauses are combined by logical conjunctions, meaning that all the clauses must be satisfied simultaneously.
The goal of the problem is to determine whether an assignment of the variables that satisfies every clause in the formula exists or not.
Similarly, the Max-$k$-SAT problem is a problem that maximizes satisfied clauses as much as possible.
When we set $k=2$, we can formulate the $2$-SAT problem with QUBO.
The $k$-SAT problem with $k\geq3$ can be formulated as HUBO directly. 

The problem Hamiltonian of the (Max-)$k$-SAT problem is given by 
\begin{align}
    \label{eq:k-sat}
    \hat{H}_\mathrm{P} = \sum_{a=1}^M\frac{1}{2^{k}}\prod_{m=1}^k(1-s_{a,m}\hat{Z}_{a,m}),
\end{align}
where $\hat{Z}_{a,m}\in\{\hat{Z}_i\}_{i=1}^N$, $s_{a,m}\in\{\pm1\}$, and $M$ is the number of clauses.
Random $k$-SAT instances are generated by selecting $k$ distinct variables uniformly for each clause and independently negating with probability $1/2$.
The hardness of the $2$-SAT and $3$-SAT problems is characterized by the ratio $\alpha=M/N$ between the number of clauses and variables~\cite{cheeseman1991really, kirkpatrick1994critical}.

\subsection{Performance comparison between the quantum and classical algorithms\label{subsec:small_benchmark}}

We use the 2-SAT problem for the performance comparison between quantum and classical algorithms.
The 2-SAT problem becomes complex around the phase transition point $\alpha\approx1$ with $N\rightarrow\infty$~\cite{cheeseman1991really, kirkpatrick1994critical}.
In this benchmark, the number of clauses is set to $M=\lfloor \alpha N\rfloor$ with clause density $\alpha=1.2$ because the critical point deviates due to the finite-size effect.
For the correspondence between the quantum and classical algorithms, we set the fixed initial state for the classical algorithms.

\begin{figure}[t]
    \centering
    \includegraphics[scale=1.0,clip]{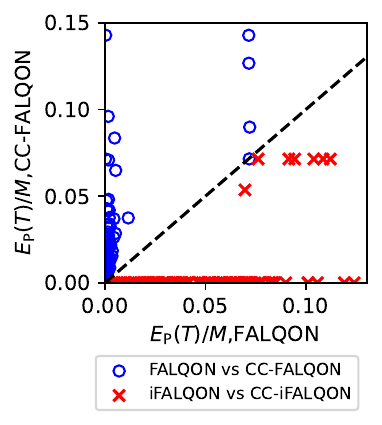}
    \caption{Energy density comparison between feedback-based quantum optimization algorithms and their classical counterparts. One hundred 2-SAT instances with $N=12$ are sampled.
    The blue circles represent the comparison between FALQON and CC-FALQON, and the red crosses represent the comparison between iFALQON and CC-iFALQON.
    The operation time is a time that satisfies the following convergence criterion: $E_\mathrm{P}(T)-E_\mathrm{P}(T+\Delta t)\leq 10^{-2}$ for $\Delta t=10^{-3}$.
    The dashed line indicates the point at which the energy density obtained by a quantum algorithm is equal to that obtained by the classical counterpart, and serves merely as a guide to the eye.}
    \label{fig:E_comparison}
\end{figure}

We first compare the quantum algorithms with their classical counterparts. 
Figure~\ref{fig:E_comparison} shows the final energy-density comparison between FALQON (iFALQON) and CC-FALQON (CC-iFALQON).
FALQON shows an advantage in solution quality compared to its classical counterpart, CC-FALQON.
The result indicates the potential quantum advantage in optimization.
On the other hand, iFALQON exhibits worse solution quality than CC-iFALQON.
That is, quantum algorithms do not always achieve an advantage compared to their classical counterparts.

Notably, the quantum and classical algorithms exhibit opposite behaviors against the inhomogeneity of driving.
In the quantum setting, homogeneous driving (FALQON) outperforms inhomogeneous driving (iFALQON).
In contrast, in the classical setting, the inhomogeneous driving (CC-iFALQON) achieves lower energy than homogeneous driving (CC-FALQON).
These different behaviors indicate the nontrivial nature of quantum and classical dynamics against the inhomogeneity of driving. 
It could be interpreted from the viewpoint of (non)locality and correlations in quantum and classical systems as discussed in Sec.~\ref{sec:discussion}.

\begin{figure}[t]
    \centering
    \includegraphics[scale=1.0,clip]{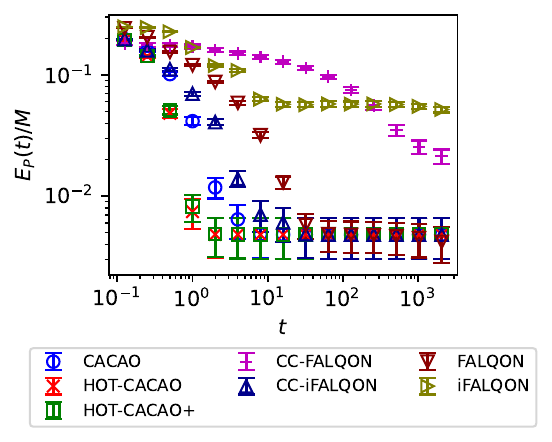}
    \caption{Energy density comparison as a function of the operation time. 
    One hundred 2-SAT instances with $N=12$ are sampled, and the error bars represent the standard deviation.
    Both the horizontal and vertical axes are on logarithmic scales.
    The blue circles represent CACAO, the red crosses represent the HOT-CACAO, the green squares represent HOT-CACAO+, the magenta plus markers represent CC-FALQON, the black blue triangles represent CC-iFALQON, the dark red triangles represent FALQON, and the olive triangles represent iFALQON.
    }
    \label{fig:T_vs_E}
\end{figure}

Next, we examine the performance of all the algorithms, i.e., FALQON, iFALQON, CC-FALQON, CC-iFALQON, CACAO, HOT-CACAO, and HOT-CACAO+.
Figure~\ref{fig:T_vs_E} shows the energy density as a function of the operation time.
FALQON, CC-iFALQON, CACAO, HOT-CACAO, and HOT-CACAO+ show similar performance in the final energy density.
Remarkably, FALQON shows slightly lower energy density than the other algorithms.
Precisely, FALQON obtains lower energy solutions in three out of one hundred instances, showing the potential to achieve quantum advantage.
In contrast, among these algorithms, classical algorithms exhibit significantly faster convergence except CC-FALQON (very slow convergence behavior of CC-FALQON is shown in Appendix~\ref{appendix:cc-falqon}).
In particular, HOT-CACAO shows the fastest convergence, followed by HOT-CACAO+.
CACAO, which incorporates only the first-order counterdiabatic term, exhibits the third fastest convergence.
This hierarchy indicates that the inclusion of higher-order counterdiabatic terms accelerates convergence, while even the first-order counterdiabatic term is effective.

Overall, these results highlight the potential of quantum algorithms and the power of classical algorithms with counterdiabatic terms. 
That is, feedback-based quantum optimization algorithms have the potential to achieve better final solution quality, suggesting an advantage in exploring the ground state, whereas the classical algorithms converge more rapidly.

\subsection{Performance evaluation of the classical algorithms for large-scale HUBO\label{subsec:3sat_banchmark}}
\begin{figure}[t]
    \centering
    \includegraphics[scale=1.0,clip]{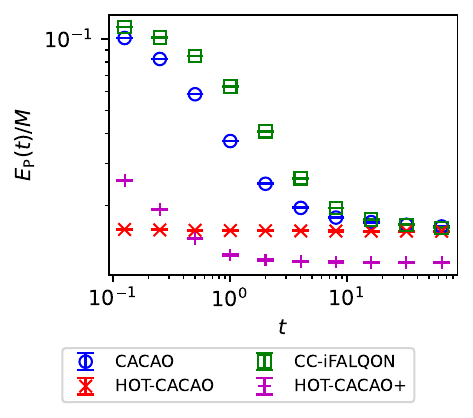}
    \caption{Energy density comparison as a function of the operation time. 
    Ten 3-SAT instances with $N=10000$ are sampled and the algorithms are performed ten times for each instance with random initial states. 
    The error bars represent the standard deviation. 
    The horizontal axis represents operation time, and the vertical axis represents energy cost. 
    Both axes are in logarithmic scale.
    The blue circles represent CACAO, the red crosses represent HOT-CACAO, the green squares represent CC-iFALQON, and the magenta plus signs represent HOT-CACAO+.
    }
    \label{fig:t_vs_E_3sat}
\end{figure}
\begin{figure}
    \centering
    \includegraphics[clip, scale=1.0]{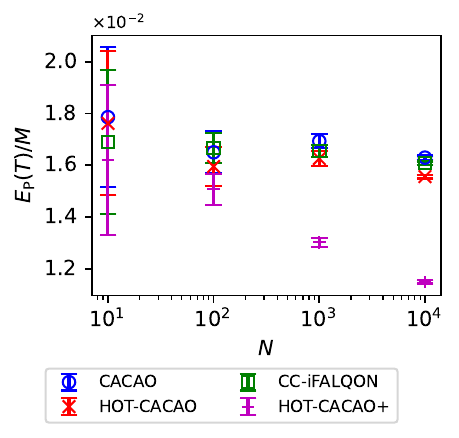}
    \caption{Energy density comparison as a function of the system size.
    Ten 3-SAT instances are sampled for each $N$ and the algorithms are performed ten times for each instance with random initial states.
    The error bars represent the standard deviation.
    The horizontal axis and the vertical axis are on a logarithmic scale.
    The blue circles represent CACAO, the red crosses represent HOT-CACAO, the green squares represent CC-iFALQON, and the magenta plus signs represent HOT-CACAO+.
    The operation time is fixed to be $T=64$.
    }
    \label{fig:energy_density_scaling_3sat}
\end{figure}
We evaluate the performance of classical algorithms for large-scale HUBO. 
Notably, classical algorithms can deal with HUBO directly without order-reduction methods.
As a benchmark model, we consider the 3-SAT problem because there is much evidence for the complexity.
The 3-SAT problem exhibits the satisfiability phase transition at $\alpha=4.26$~\cite{cheeseman1991really, kirkpatrick1994critical}. 
In this study, we set the number of clauses as $M=\lfloor \alpha N\rfloor$ with the ratio $\alpha=4.2$ because of the finite-size effect~\cite{barthel2002hiding}.
In this benchmark, we obtain the average energy density over ten different random initial states generated from uniform random as mentioned in Sec~\ref{subsec:settings_classical}.
The importance of the sampling on random initial states is shown in Appendix.~\ref{appendix:small_properties}. 

We apply classical algorithms to large-scale random 3-SAT instances with the system size $N=10000$.
Based on the results in the previous section, we focus on representative classical algorithms that exhibit relatively fast convergence, i.e., CC-iFALQON, CACAO, HOT-CACAO, and HOT-CACAO+. 
Figure~\ref{fig:t_vs_E_3sat} shows the energy density as a function of the operation time for these algorithms.
These classical algorithms achieve very low energy density. 
In particular, HOT-CACAO+ achieves the lowest final energy among all the methods considered.
Since each unsatisfied clause contributes the energy of $1$ (while satisfied clauses contribute $0$), energy density on the order of $10^{-2}$ indicates that only a small fraction of clauses remains unsatisfied.
This result indicates that the inclusion of both higher-order counterdiabatic terms and the problem Hamiltonian is important to combat the difficulty of the 3-SAT problem, incorporating higher-order interactions for managing three-body interactions of the 3-SAT problem and the grobal structure of the problem explicitly.

The improvement of solution quality with the inclusion of the problem Hamiltonian, however, comes at the cost of slower convergence. 
That is, HOT-CACAO+ shows slower convergence than HOT-CACAO, while HOT-CACAO+ reaches lower energy than HOT-CACAO. 
A similar trend is observed for CC-iFALQON against CACAO; the former includes the problem Hamiltonian in addition to the local terms and the latter only consists of the local terms.
These observations support the above statement.

We further investigate the scalability of these methods by examining the energy density as a function of the system size. 
Figure~\ref{fig:energy_density_scaling_3sat} shows the energy density as a function of the system size.
We find that the performance of CACAO, HOT-CACAO, and CC-iFALQON are almost constant as the problem size increases.
In contrast, HOT-CACAO+ even achieves lower energy density for larger system size.
This result demonstrates the superior scalability of HOT-CACAO+ and supports its effectiveness for further large-scale optimization problems.

Overall, these findings indicate that, while the inclusion of only counterdiabatic terms provides faster convergence, incorporating both higher-order counterdiabatic terms and the problem Hamiltonian is crucial for achieving high-quality solutions in large-scale, high-order optimization problems.
Also, these proposed algorithms have the potential as direct HUBO solvers.

\section{discussion\label{sec:discussion}}
In this section, we discuss the results obtained by the benchmark tests and perform a further detailed analysis.
We first consider opposite behaviors of the quantum and classical algorithms with respect to the inhomogeneity of driving. 
As a possible interpretation, we focus on how the nature of dynamics---such as (non)locality and correlations---affects both the convergence properties and the quality of achievable solutions.
To this end, we provide a unified explanation for the contrasting performance of the quantum and classical algorithms.
Next, we analyze the dynamical properties of classical algorithms in more detail, clarifying how the inclusion of higher-order counterdiabatic terms and the problem Hamiltonian influences the convergence behavior.

\subsection{(Non)locality and correlations in quantum and classical algorithms \label{dis:locality_correration}}
The different effects of the inhomogeneity in the quantum and classical algorithms could be attributed to the nature of quantum and classical dynamics, i.e., (non)locality of dynamics.

Quantum dynamics is determined globally by the Schr\"{o}dinger equation~\eqref{eq:Seq}.
The parameter of FALQON in Eq.~\eqref{eq:falqon_param} globally accumulates the infomation of the problem Hamiltonian. 
In contrast, the parameters of iFALQON are determined rather locally for each spin (indeed, $[\hat{H}_\mathrm{P},\hat{X}_i]$ is a local operator).
This viewpoint could explain why homogeneous control in FALQON yields better results than inhomogeneous control in iFALQON, i.e., the mismatch between global dynamics and local parameters would degrade the performance.

We then extend the above discussion to classical algorithms. 
In classical algorithms, dynamics is governed by local updates as described in Eq.~\eqref{eq:equation_of_motion}, i.e., classical dynamics is inherently local. 
Thus, classical algorithms can benefit from inhomogeneous driving, which enables an independent improvement in each degree of freedom.
In contrast, homogeneous driving is generally mismatched with the underlying local dynamics, leading to limited performance.

Moreover, better solution quality of FALQON and faster convergence of classical algorithms could be attributed to the presence and absence of nonlocal correlations in quantum and classical dynamics. 
That is, FALQON can globally explore the solution space with global correlations, while classical algorithms simplifies the optimization landscape and allows more direct descent toward low-energy states.

Overall, we give a possible mechanism for the contrasting behaviors of quantum and classical algorithms. 
That is, global nature of quantum dynamics and local nature of classical dynamics cause the different, contrasting behaviors.

\subsection{Dynamics of classical algorithms\label{dis:classical_algorithms}}
\begin{figure}[t]
    \centering
    \includegraphics[clip, scale=1.0]{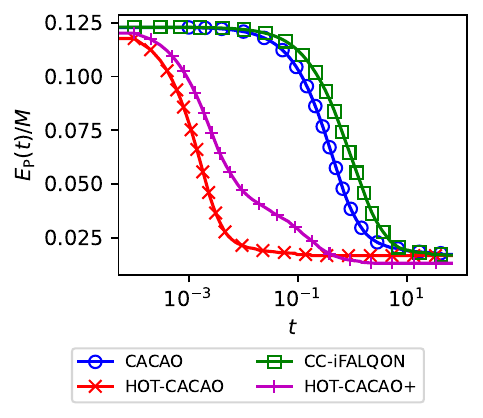}
    \caption{Energy density dynamics of classical algorithms applied to a 3-SAT instance with $N=1000$.
    The algorithms are performed with a random initial state. 
    The horizontal axis represents the operation time $t$ on a logarithmic scale, and the vertical axis represents the energy density.
    The blue circles represent CACAO, the red crosses represent the HOT-CCAO, the green squares represent CC-iFALQON, and the magenta plus markers represent HOT-CACAO+.
    Solid lines between points are provided as a guide to the eye.
    For clarity, only 20 data points with respect to the operation time are plotted.
    }
    \label{fig:energy_dynamics_3sat}
\end{figure}
\begin{figure}[t]
    \centering
    \includegraphics[clip, scale=1.0]{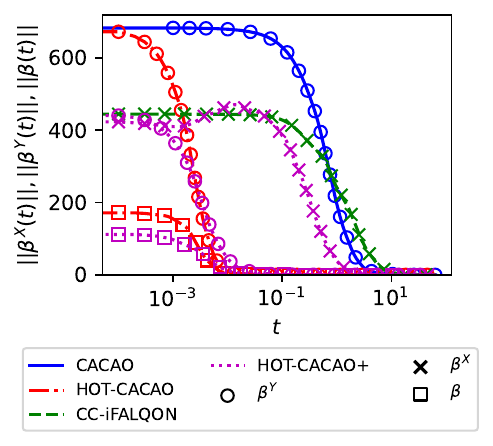}
    \caption{Control strengths of classical algorithms for a 3-SAT instance with $N=1000$ as a function of the operation time.
    The algorithms are performed with a random initial state. 
    The horizontal axis represents the operation time $t$ on a logarithmic scale, and the vertical axis represents the control strengths.
    The blue (solid), red (dash-dotted), green (dashed), and magenta (dotted) plots (lines) represent the control strengths for CACAO, HOT-CACAO, CC-iFALQON, and HOT-CACAO+, respectively.
    The circles, crosses, and squares represent $\|\beta^Y\|$, $\|\beta^X\|$, and $\|\beta\|$, respectively.
    Solid lines between points are provided as a guide to the eye.
    For clarity, only 20 data points with respect to the operation time are plotted.
    }
    \label{fig:control_strength}
\end{figure}
In this section, we examine the dynamical properties of classical algorithms by considering a 3-SAT instance with $ N=1000$.
Figure~\ref{fig:energy_dynamics_3sat} shows the energy dynamics for CACAO, HOT-CACAO, CC-iFALQON, and HOT-CACAO+.
It is observed that CACAO and CC-iFALQON exhibit similar dynamical behavior.
In contrast, HOT-CACAO and HOT-CACAO+ demonstrate significantly faster convergence compared to these algorithms.
While HOT-CACAO converges to approximately the same final energy as CACAO and CC-iFALQON, HOT-CACAO+ reaches lower energy states.

To investigate the origin of this behavior, we analyze the strength of the control parameters.
The following norm gives the control strengths
\begin{align}
    \label{eq:control_strength}
    &\|\beta^W(t)\| = \sum_{i=1}^N|\beta_i^W(t)|,\quad W=X,Y,\nonumber\\
    &\|\beta(t)\| = \sum_{\substack{i,j=1 \\ (i\neq j)}}^N|\beta_{ij}(t)|.
\end{align}
The control strengths $\|\beta^W\|$ and $\|\beta\|$ reflect the behavior of dynamics because these are state-dependent quantities. 

Figure~\ref{fig:control_strength} shows the control strengths as a function of the operation time.
CACAO and CC-iFALCON employ only first-order control terms, i.e., they are governed solely by $\{\beta_i^Y(t)\}$ and $\{\beta_i^X(t)\}$, respectively.
In this first-order setting, we observe relatively slow convergence.
In contrast, when second-order terms are introduced in HOT-CACAO, the control strengths converge more rapidly, even for the first-order terms, indicating enhanced dynamical efficiency.
In this setting, the behavior of the first-order terms follows the behavior of the second-order terms. 
HOT-CACAO+ incorporates multiple control terms, where $\|\beta^Y(t)\|$ and $\|\beta(t)\|$ exhibit rapid convergence similar to HOT-CACAO, whereas $\|\beta^X(t)\|$ converges on a timescale comparable to that of CACAO and CC-iFALQON.
This behavior suggests that the dynamics in the $X$-direction relaxes quickly, while exploration in the $Y$--$Z$ plane---governed by terms $\{\beta_i^X(t)\}$---continues over a longer timescale.

In HOT-CACAO+, the nontrivial, non-monotonic behavior of $\|\beta^X(t)\|$, which exhibits a pronounced peak during the evolution, is also observed. 
It may enable more effective exploration of the energy landscape, allowing the system to reach lower-energy states than other methods.
These observations suggest that increasing the degrees of freedom in the control parameters plays a crucial role in improving optimization performance.
Note that, in small-scale problems, the behavior becomes more chaotic (see Appendix~\ref{appendix:small_properties}).

\section{Conclusion\label{sec:conclusion}}
In this paper, we have introduced the classical counterpart of feedback-based quantum optimization by using the quantum-classical correspondence of spin systems, and investigated the performance of the quantum and classical algorithms.
Through systematic comparisons, we have clarified the fundamental differences between quantum and classical dynamics in terms of solution quality, convergence behavior, and scalability.

We have first shown that FALQON can achieve lower energy than its classical counterpart and other classical algorithms, indicating its advantage in exploring the global structure of the optimization landscape. 
In contrast, some classical algorithms have exhibited faster convergence than FALQON and similar solution quality.
Among them, HOT-CACAO and HOT-CACAO+ have demonstrated rapid convergence, highlighting the effectiveness of the second-order counterdiabatic term in accelerating optimization dynamics.

We have also applied the classical algorithms to the 3-SAT problem, which includes three-body interactions.
In large-scale problems with $N=10000$, we have demonstrated that HOT-CACAO+ achieves the best performance among the classical algorithms. 
This result indicates that combining higher-order counterdiabatic terms with the explicit inclusion of the problem Hamiltonian $\mathcal{H}_\mathrm{P}$ is essential to achieve lower-energy solutions in complex optimization problems.
However, this improvement has come with relatively slower convergence than HOT-CACAO, revealing a trade-off between convergence speed and solution quality. 

Finally, scalability analysis has showed that only HOT-CACAO+ consistently improves the energy density as the system size increases, whereas other methods keep the performance almost constant in larger systems. 
This again highlights the importance of incorporating both higher-order counterdiabatic terms and the problem Hamiltonian for achieving scalable optimization.
The scalability analysis suggests that classical algorithms have the potential to become promising solvers for HUBO problems.

Overall, our results have demonstrated the potential of both feedback-based quantum optimization algorithms and classical algorithms.
FALQON has the capability to achieve high-quality solutions, while classical algorithms avoid iterative overhead and can obtain near-optimal solutions with fast convergence.

It remains an open question which method achieves the best overall performance under given restrictions, considering the tradeoff between FALQON and classical algorithms and that between HOT-CACAO and HOT-CACAO+. 
Thus, systematic comparisons, for example, under equal computational budgets, is left to the future work.
Nevertheless, classical algorithms seem more efficient than quantum algorithms, as they do not require access to intermediate quantum states and the implementation of iterative processes.

Considering the direct higher-order formulation is the best approach for solving HUBO~\cite{ikeuchi2025evaluating}, classical algorithms would also be advantageous because we can directly deal with higher-order terms without order-reduction methods.
The application of classical algorithms to a wide range of combinatorial optimization problems~\cite{micheletti2021polymer, chermoshentsev2021polynomial, outeiral2021investigating, slongo2023quantum, jun2023hubo}, and systematic comparisons with quantum-inspired optimization solvers~\cite{yamaoka201520k, inagaki2016coherent, fixstars_amplify, goto2019combinatorial, goto2021high, goto2025edge} remain important directions for the future work.

\begin{acknowledgments}
TH1 was supported by JST SPRING (Grant Number JPMJSP2123).
This work was supported by JST Moonshot R\&D Grant No. JPMJMS2061.
\end{acknowledgments}

\appendix
\section{Quantum-classical correspondence of spin systems and the canonical equations of motion for classical spins\label {appendix:quantum_classical_correspondence}}

Quantum spin systems consisting of the Pauli matrices $\{\hat{X}_i,\hat{Y}_i,\hat{Z}_i\}_{i=1}^N$ are characterized by the commutation relations
\begin{equation}
\begin{aligned}
&[\hat{X}_i,\hat{Y}_i]=2\mathrm{i}\hat{Z}_i,\\
&[\hat{Y}_i,\hat{Z}_i]=2\mathrm{i}\hat{X}_i,\\
&[\hat{Z}_i,\hat{X}_i]=2\mathrm{i}\hat{Y}_i,
\end{aligned}
\end{equation}
and otherwise zeros. 
The quantum-classical correspondence requires the following commutation relations
\begin{equation}
\begin{aligned}
&\{m_i^X,m_i^Y\}=2m_i^Z,\\
&\{m_i^Y,m_i^Z\}=2m_i^X,\\
&\{m_i^Z,m_i^X\}=2m_i^Y,
\end{aligned}
\end{equation}
and otherwise zeros for the classical spins $\{m_i^X,m_i^Y,m_i^Z\}_{i=1}^N$, where the bracket is the Poisson bracket $\{\bullet,\star\}=\sum_{i=1}^N(\frac{\partial\bullet}{\partial q_i}\frac{\partial\star}{\partial p_i}-\frac{\partial\star}{\partial q_i}\frac{\partial\bullet}{\partial p_i})$ with generalized coordinates $\{q_i,p_i\}_{i=1}^N$. 
We can show that the classical spins satisfy the above commutation relations if the generalized coordinates are defined as
\begin{equation}\label{eq:canonical_variables}
\begin{aligned}
&m_i^X=\sqrt{1-(2p_i)^2}\cos{q_i},\\
&m_i^Y=\sqrt{1-(2p_i)^2}\sin{q_i},\\
&m_i^Z=2p_i. 
\end{aligned}
\end{equation}

As quantum dynamics is governed by the Schr\"odinger equation (\ref{eq:Seq}), classical dynamics is governed by the Hamilton equation
\begin{equation}
\label{eq:Hamilton}
\dot{q}_i = \frac{\partial \mathcal{H}_t}{\partial p_i},\quad
\dot{p}_i = -\frac{\partial \mathcal{H}_t}{\partial q_i}.
\end{equation}
We can show that the time derivative of Eq.~(\ref{eq:canonical_variables}) with the Hamilton equation (\ref{eq:Hamilton}) gives the canonical equations of motion (\ref{eq:equation_of_motion}). 
It can be immediately confirmed by using
\begin{equation}
\begin{aligned}
&\frac{\partial\mathcal{H}_t}{\partial p_i}=\frac{\partial m_i^X}{\partial p_i}\frac{\partial\mathcal{H}_t}{\partial m_i^X}+\frac{\partial m_i^Y}{\partial p_i}\frac{\partial\mathcal{H}_t}{\partial m_i^Y}+\frac{\partial m_i^Z}{\partial p_i}\frac{\partial\mathcal{H}_t}{\partial m_i^Z},\\
&\frac{\partial\mathcal{H}_t}{\partial q_i}=\frac{\partial m_i^X}{\partial q_i}\frac{\partial\mathcal{H}_t}{\partial m_i^X}+\frac{\partial m_i^Y}{\partial q_i}\frac{\partial\mathcal{H}_t}{\partial m_i^Y}.
\end{aligned}
\end{equation}
We can also check the quantum-classical correspondence of spin dynamics by applying the classical (mean-field) approximation to the Heisenberg equation of quantum spins.

\section{Convergence behavior of CC-FALQON\label{appendix:cc-falqon}}
\begin{figure}[t]
    \centering
    \includegraphics[scale=1.0,clip]{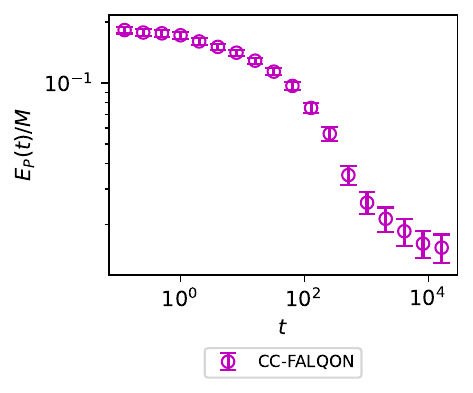}
    \caption{Energy density of CC-FALQON as a function of the operation time. 
    One hundred 2-SAT instances with $N=12$ are sampled, and the error bars represent the standard deviation.
    Both the horizontal and vertical axes are on logarithmic scales.
    }
    \label{fig:T_vs_E_ccfalqon}
\end{figure}
\begin{figure}[t]
    \centering
    \includegraphics[scale=1.0,clip]{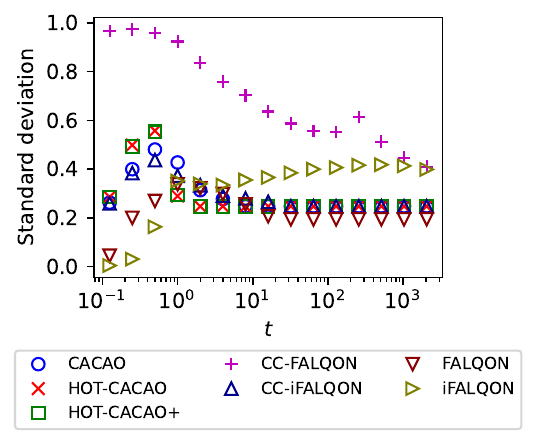}
    \caption{Standard deviation of the energy density of all the algorithms as a function of the operation time. 
    One hundred 2-SAT instances with $N=12$ are sampled.
    Both the horizontal axis is on a logarithmic scale.
    The blue circles represent CACAO, the red crosses represent the HOT-CACAO, the green squares represent HOT-CACAO+, the magenta plus markers represent CC-FALQON, the black blue triangles represent CC-iFALQON, the dark red triangles represent FALQON, and the olive triangles represent iFALQON. 
    }
    \label{fig:T_vs_variance}
\end{figure}
In Sec.~\ref{subsec:small_benchmark}, we have examined the convergence behavior of the algorithms for random 2-SAT instances.
Among them, CC-FALQON and iFALQON have exhibited very slow convergence. 
Here, we check the convergence behavior of CC-FALQON for a long timescale as a representative example. 
Figure~\ref{fig:T_vs_E_ccfalqon} presents the energy density of CC-FALQOON as a function of the operation time.
It is obvious that CC-FALQON requires a longer time than $10^{4}$ for convergence.

Figure~\ref{fig:T_vs_variance} presents the standard deviation of the energy density of all the algorithms as a function of the operation time.
Compared to the other algorithms, CC-FALQON exhibits a substantially larger standard deviation in the energy density.
This result suggests that CC-FALQON cannot systematically reduce the energy density. 
Thus, CC-FALQON is impractical and unreliable compared with the other algorithms considered.

\section{Importance of random initial state sampling\label{appendix:small_properties}}
\begin{figure*}[t]
    \centering
    \includegraphics[clip, scale=1.0]{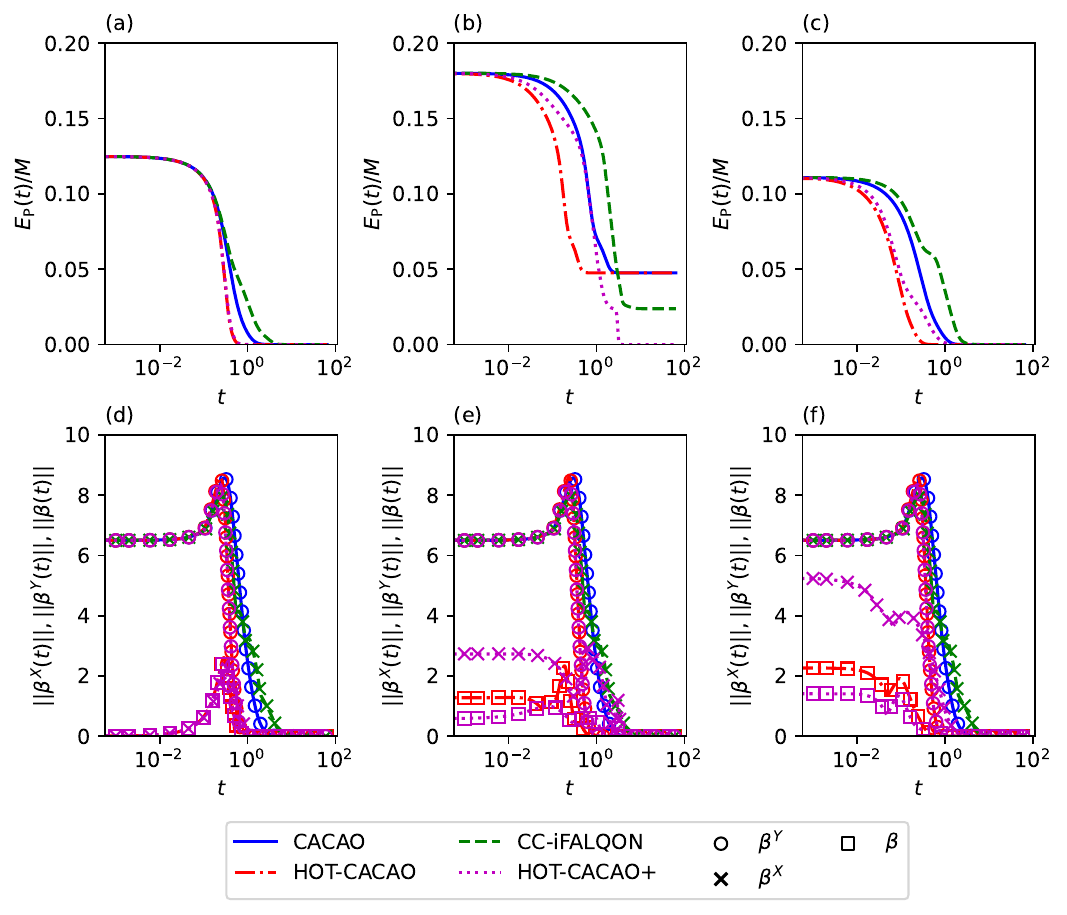}
    \caption{Energy density dynamics and control strength of classical algorithms as a function of the operation time. 
    A 3-SAT instance with $N=10$ is considered for the fixed initial state [(a) and (d)] and two random initial states [(b) and (e), and (c) and (f)].
    The horizontal axis represents the operation time, and the vertical axes represent the energy density [(a), (b), and (c)] and control strength [(d), (e), and (f)], respectively.
    The blue (solid), red (dash-dotted), green (dashed), and magenta (dotted) plots (lines) represent control strengths for CACAO, HOT-CACAO, CC-iFALQON and HOT-CACAO+, respectively.
    The circles, crosses and squares represent $\|\beta^Y\|$, $\|\beta^Y\|$ and $\|\beta\|$, respectively.
    Solid lines between points are provided as a guide to the eye.
    For clarity, only 20 data points with respect to the operation time are plotted.
    }
    \label{fig:small_scale_3sat_result}
\end{figure*}
We point out the importance of random initial state sampling because of chaotic behavior during optimization. 
We consider the application of classical algorithms to a 3-SAT instance with $N=10$.
Figures~\ref{fig:small_scale_3sat_result}~(a) and~(d) show the energy dynamics and control parameter strengths obtained from the fixed initial state, while Figs.~\ref{fig:small_scale_3sat_result}~(b) and~(e), and (c) and~(f) correspond to different random initial states for the same 3-SAT instance.
Comparing these results reveals that both the final energy and the relative performance of the algorithms depend strongly on the choice of the initial condition.
Thus, random initial state sampling is important to suppress the bad influence of chaotic behavior and rather to benefit from it.

\bibliography{reference}
\end{document}